\RequirePackage[2020-02-02]{latexrelease}
\documentclass[aps,floatfix,twocolumn]{revtex4}
\usepackage{graphicx}
\usepackage{dcolumn}
\usepackage{bm}
\usepackage{float}
\usepackage[verbose]{placeins}
\usepackage{amsmath}
\usepackage{romannum}
\usepackage{csquotes}
\usepackage[breaklinks=true,colorlinks=true,linkcolor=black,urlcolor=black,citecolor=black]{hyperref}
\hypersetup{
    colorlinks = true,
    citecolor = blue,
    linkcolor = blue,
     urlcolor = blue,
}      
\begin{document}
\title{Revisiting thermoelectric transport across strongly correlated quantum dot: A Green's function equation of motion theory perspective}
\author{\bf{Sachin Verma$^1$}}
\author{\bf{B. S. Tewari$^2$}}
\author{\bf{A. Dhyani$^3$}}
\author{\bf{Ajay$^1$}}
\email{$^1$sverma2@ph.iitr.ac.in\\$^2$bhagyasindhu@gmail.com\\
$^3$archana.d2003@gmail.com\\
$^1$ajay@ph.iitr.ac.in}
\affiliation{$^1$Department of Physics, Indian Institute of Technology Roorkee, 247667, Uttarakhand, India}
\affiliation{$^2$Department of Applied Sciences and Humanities, G. B. Pant Institute of Engg. and Tech., Pauri Garhwal, 246194, Uttarakhand, India}
\affiliation{$^3$Department of Applied Sciences, School of Engg., University of Petroleum and Energy Studies, Dehradun, 248007, Uttarakhand, India}

\begin{abstract}
Using Green's function equation of motion within Lacroix decoupling scheme, we examine the thermoelectric transport features of a strongly interacting quantum dot coupled between metallic leads. We demonstrate that a qualitative description of the thermoelectric transport in the Kondo regime requires a complete self-consistent treatment of Green's function. The linear thermoelectric properties, including electrical conductivity, thermal conductivity, thermopower, and figure of merit, are analyzed as a function of temperature ranging from Kondo to Coulomb blockade regime. The results presented here are qualitatively consistent with existing results obtained using different theoretical techniques.
\end{abstract}

\maketitle
\pagenumbering{arabic}

\section{Introduction}
The study of thermoelectric transport across low-dimensional systems has been a topic of significant interest in the past three decades due to its potential applications in nanoscale thermoelectric devices\cite{Dubi2011,Sothmann2014,Pekola2021,Hicks1993-96,Mahan1996}. Particularly, the thermoelectric transport in a hybrid systems that combines metallic leads and quantum dots has attracted a lot of interest both theoretically\cite{Beenakker1992,Andreev2001,Humphrey2002,Turek2002,Koch2004,Humphrey2005,Zianni2007,Kubala2008,Murphy2008,Franco2008,Nakpathomkun2010,Mani2011,Kennes2013,Taylor2015,Erdman2017,Manaparambil2023,Boese2001, Krawiec2006,Krawiec2007,Dong2002,Kubala2006,Costi1994,Costi2010} and experimentally\cite{Svilans2016,Staring1993,Dzurak1997,Appleyard2000,Small2003,Scheibner2005,Scheibner2007,Svensson2012,Svensson2013,Josefsson2018,Jaliel2019,Dutta2019,Dorsch2021,Svilans2018,Erdman2019}. These hybrid quantum dot(QD) systems can also be used as a testing ground for various theoretical techniques for further analyzing the transport properties of novel strongly correlated materials.\\
Due to the Seebeck effect, the hybrid QD system works as a particle-exchange heat engine in the presence of a thermal gradient across the source and drain reservoir\cite{Humphrey2002}. These hybrid QD heat engines are prone to evolve interesting thermoelectric transport features governed by quantum confinement, the Coulomb blockade effect, and the Kondo effect on QD. Due to quantum confinement and Coulomb blockade effects, these QD heat engines are more efficient at converting thermal energy into electrical energy than their bulk counterparts\cite{Hicks1993-96,Mahan1996}. The enhancement of the thermoelectric efficiency  is caused by a strong violation of the Wiedemann–Franz law\cite{Kubala2008} and the significant drop in lattice thermal conductivity\cite{Balandin1998, Tsaousidou2010}. Although there have been several theoretical and experimental studies on QD heat engines in various parameter regimes, we will only concentrate on the thermoelectric transport in Kondo-correlated QD in a linear response regime, i.e., for small voltage or thermal biasing.\\
A Strong Coulomb interaction on the QD energy level results in an unpaired electron with localized spin. At very low temperatures ($T\leq T_K$, where $T_K$ is Kondo temperature) and with strong QD reservoir coupling, the localized QD spin can interact with the conduction electrons of reservoirs, leading to the formation of a many-body singlet state. This many-body state manifests as a sharp Kondo resonance in the conductance and can dramatically alter the transport properties of the system\cite{Gordon1998, Cronenwett1998, Kouwenhoven22001, Pustilnik2004}. The thermoelectric effect in Kondo-correlated QD has been studied in detail in earlier research works using various theoretical techniques such as the non-perturbative resonant tunneling approximation\cite{Boese2001}, perturbation theory\cite{Dong2002,Kubala2006,Kubala2008}, Slave-Boson or modiefied Slave-Boson method \cite{Krawiec2006,Krawiec2007,Franco2008}, and more rigorous numerical renormalization-group(NRG) technique\cite{Costi1994,Costi2010,Bas2012,Weymann2013,Quintero2021}. However, the self-consistent Green's function equation of motion (EOM) technique has not been tested adequately in investigating the thermoelectric transport properties of the hybrid QD system in the Kondo regime. Therefore, we revisit the fundamental mechanisms underlying the thermoelectric effects in strongly interacting QD-based devices for temperatures ranging from Coulomb blockade to the Kondo regime.\\
The present paper aims to provide a pedagogical approach using Green's function EOM method that elucidates the intricate interplay between thermoelectric and Kondo physics. We have also compared two approximate solutions of the QD Green's function in the EOM technique widely used to analyze the Kondo effect in QD systems. In the next section, we begin with a thorough derivation of the appropriate Green's functions. We then present the result and discussion of the linear thermoelectric transport properties, highlighting the unique features that arise in the Coulomb blockade and  Kondo regime.
\section{Theoretical formalism}\label{sec:artwork}
The normal metal-quantum dot system is modelled by second quantized single impurity Anderson Hamiltonian\cite{Anderson1961},
\begin{equation} \label{eq:1}
\begin{aligned}
H = & \sum_{k\sigma,\alpha}\epsilon_{k,\alpha}c^\dagger_{k\sigma,\alpha}c_{k\sigma,\alpha}+ \sum_{\sigma} \epsilon_{d}d^\dagger_{\sigma}d_{\sigma}+U n_{\uparrow} n_{\downarrow}+\\
& \sum_{k\sigma,\alpha}(V_{k,\alpha}d^\dagger_{\sigma}c_{k\sigma,\alpha}+V^{\ast}_{k,\alpha}c^\dagger_{k\sigma,\alpha}d_{\sigma})]
\end{aligned}
\end{equation}
First term represents the Hamiltonian of the free electronic band, where $\epsilon_{k,\alpha}$ is the kinetic energy of the electrons in the $\alpha\in L,R$ reservoir, and $c_{k\sigma,\alpha}$($c^\dagger_{k\sigma,\alpha }$) is the annihilation (creation) operator of an electron with spin $\sigma$ and wave vector $\vec{k}$. Second term represents the Hamiltonian of the quantum dot with $n_{\sigma}=d_\sigma^\dagger d_\sigma$ as the number operator and  $d_\sigma(d^\dagger_\sigma)$ is the annihilation (creation) operator of electron with spin $\sigma$. The quantum dot consists of a single electronic level of energy $\epsilon_d$ and can be occupied by up to two electrons, i.e., $\sigma\in\uparrow,\downarrow$. The third term represents the on-dot electron-electron Coulomb repulsion with the interaction strength $U$. Last term represents the tunneling Hamiltonian or hybridization between the QD and $\alpha$ reservoirs with coupling strength $V_{k}$.\\
The Green's function equation of motion (EOM) method in Fourier space with Zubarev notation\cite{Zubarev1960} is employed to address effective model Hamiltonian in Eqn. (1). As we are interested in physical processes occurring in some small energy range around the Fermi level, the tunneling amplitude is considered $k$ independent, i.e., $V_{k,\alpha}=V_{\alpha}$ for $V_{k,\alpha}<<D$ (wide band), where $-D\leq\epsilon_{k,\alpha}\leq D$, with $D$ as the half bandwidth. The tunneling coupling strength of the QD to the $\alpha$ reservoir is defined by $\Gamma_{\alpha}=2\pi|V_{\alpha}|^2\rho_{0\alpha}$, where metallic density of states $\rho_{0\alpha}$ is constant in the range of energy around Fermi level (flat band).\\
The Fourier transform of the single particle retarded Green’s function or propagator for QD
\begin{align}
G_{d\sigma}^r(\omega)&=\langle\langle{d_{\sigma}|d^\dagger_{\sigma}}\rangle\rangle\nonumber\\
&=\cfrac{-i}{2\pi}\lim_{\delta \to 0^{+}}\int{\theta(t)\langle\{d_{\sigma}(t),d^\dagger_{\sigma}(0)\}\rangle e^{i(\omega+i\delta)t}}dt
\end{align}
with spin $\sigma$ must satisfies the following EOM
\begin{equation} \label{eq:2}
\omega\langle\langle{d_{\sigma}|d^\dagger_{\sigma}}\rangle\rangle=\langle\{d_{\sigma},d^\dagger_{\sigma}\}\rangle+\langle\langle[d_{\sigma},H]|d^\dagger_{\sigma}\rangle\rangle.
\end{equation}
By evaluating the commutator and anti-commutator brackets in the preceding equation for a single particle retarded Green's function of QD,
\begin{equation} \label{eq:3}
\left(\omega-\epsilon_{d}-\sum_{k\alpha}\frac{|V_{\alpha}|^2}{(\omega-\epsilon_{k,\alpha})}\right)\langle\langle{d_{\sigma}|d^\dagger_{\sigma}}\rangle\rangle=1+U\langle\langle{d_{\sigma}n_{\bar{\sigma}}|d^\dagger_{\sigma}}\rangle\rangle
\end{equation}
Eqn. (4) shows that EOM for a given Green's function with a non-zero Coulomb interaction requires higher-order correlation functions. The equation of motion for higher-order correlation functions is given as
\begin{widetext}
\begin{align} 
\label{eq:4}&\nonumber\left(\omega-\epsilon_{d}-U\right)\langle\langle{d_{\sigma}n_{\bar{\sigma}}|d^\dagger_{\sigma}}\rangle\rangle                                              =\langle{n_{\bar{\sigma}}}\rangle+\sum_{k,\alpha}V^{\ast}_{\alpha}\langle\langle{c_{k\sigma,\alpha}n_{\bar{\sigma}}|d^\dagger_{\sigma}}\rangle\rangle-\sum_{k,\alpha}V^{\ast}_{\alpha}\langle\langle{c_{k\bar{\sigma},\alpha}d_{\bar{\sigma}}^{\dagger}d_{\sigma}|d^\dagger_{\sigma}}\rangle\rangle+\\
&\hspace{32.7em}\sum_{k,\alpha}V^{\ast}_{\alpha}\langle\langle{c_{k\bar{\sigma},\alpha}^{\dagger}d_{\sigma}d_{\bar{\sigma}}|d^\dagger_{\sigma}}\rangle\rangle\\  
\nonumber\\[2pt]
\label{eq:5}&\left(\omega-\epsilon_{k,\alpha}\right)\langle\langle{c_{k\sigma,\alpha}n_{\bar{\sigma}}|d^\dagger_{\sigma}}\rangle\rangle = V_{\alpha}\langle\langle{d_{\sigma}n_{\bar{\sigma}}|d^\dagger_{\sigma}}\rangle\rangle+\sum_{k^{'}\alpha}V_{\alpha}^{\ast}\langle\langle{c_{k^{'}\bar{\sigma},\alpha}^{\dagger}c_{k\sigma,\alpha}d_{\bar{\sigma}}|d_{\sigma}^{\dagger}}\rangle\rangle-\sum_{k^{'}\alpha}V_{\alpha}\langle\langle{c_{k\sigma,\alpha}c_{k^{'}\bar{\sigma},\alpha}d_{\bar{\sigma}}^{\dagger}|d_{\sigma}^{\dagger}}\rangle\rangle\\
\nonumber\\[2pt]
\label{eq:6}&\left(\omega-\epsilon_{k,\alpha}\right)\langle\langle{c_{k\bar{\sigma},\alpha}d_{\bar{\sigma}}^{\dagger}d_{\sigma}|d^\dagger_{\sigma}}\rangle\rangle = -\langle{d_{\bar{\sigma}}^{\dagger}c_{k\bar{\sigma},\alpha}}\rangle-V_{\alpha}\langle\langle{d_{\sigma}n_{d\bar{\sigma}}|d^\dagger_{\sigma}}\rangle\rangle+\sum_{k^{'}\alpha}V_{\alpha}\langle\langle{c_{k^{'}\bar{\sigma},\alpha}^{\dagger}c_{k\bar{\sigma},\alpha}d_{\sigma}|d_{\sigma}^{\dagger}}\rangle\rangle-\nonumber\\
&\hspace{31.5em}\sum_{k^{'}\alpha}V_{\alpha}^{\ast}\langle\langle{c_{k\sigma,\alpha}c_{k^{'}\bar{\sigma},\alpha}d_{\bar{\sigma}}^{\dagger}|d_{\sigma}^{\dagger}}\rangle\rangle\\
\nonumber\\[2pt]
\label{eq:7}&\left(\omega+\epsilon_{k,\alpha}-2\epsilon_d-U\right)\langle\langle{c_{k\bar{\sigma},\alpha}^{\dagger}d_{\sigma}d_{\bar{\sigma}}|d^\dagger_{\sigma}}\rangle\rangle = -\langle{c_{k\bar{\sigma},\alpha}^{\dagger}d_{\bar{\sigma}}}\rangle+V_{\alpha}^{\ast}\langle\langle{d_{\sigma}n_{\bar{\sigma}}|d^\dagger_{\sigma}}\rangle\rangle-\sum_{k^{'}\alpha}V_{\alpha}^{\ast}\langle\langle{c_{k^{'}\bar{\sigma},\alpha}^{\dagger}c_{k\bar{\sigma},\alpha}d_{\sigma}|d_{\sigma}^{\dagger}}\rangle\rangle+\nonumber\\
&\hspace{31.5em}\sum_{k^{'}\alpha}V_{\alpha}^{\ast}\langle\langle{c_{k^{'}\bar{\sigma},\alpha}^{\dagger}c_{k\sigma,\alpha}d_{\bar{\sigma}}|d_{\sigma}^{\dagger}}\rangle\rangle 
\end{align}
\end{widetext}
Again, these Eqns. (5)-(8) contain higher order correlation functions, resulting in a hierarchy of equations. Therefore, a decoupling scheme is required to truncate this hierarchy of equations. In the present work, we study thermoelectric transport in the Coulomb blockade and Kondo regimes at very strong on-dot Coulomb repulsion ($U\rightarrow\infty$) using a decoupling scheme beyond the mean-field approximation. Therefore, we close the system of equations by using Lacroix decoupling scheme\cite{Lacroix1981,Meir1991,Wohlman2005,Fang2008} i.e. 
\begin{align*}
&\langle\langle{c_{k\sigma,\alpha}c_{k^{'}\bar{\sigma},\alpha}d_{\bar{\sigma}}^{\dagger}|d_{\sigma}^{\dagger}}\rangle\rangle \approx -\langle{d_{\bar{\sigma}}^{\dagger}c_{k^{'}\bar{\sigma},\alpha}}\rangle \langle\langle{c_{k\sigma,\alpha}|d_{\sigma}^{\dagger}}\rangle\rangle\\[10pt]
&\langle\langle{c_{k^{'}\bar{\sigma},\alpha}^{\dagger}c_{k\sigma,\alpha}d_{\bar{\sigma}}|d_{\sigma}^{\dagger}}\rangle\rangle \approx -\langle{c_{k^{'}\bar{\sigma},\alpha}^{\dagger}d_{\bar{\sigma}}}\rangle \langle\langle{c_{k\sigma,\alpha}|d_{\sigma}^{\dagger}}\rangle\rangle\\[10pt] &\langle\langle{c_{k^{'}\bar{\sigma},\alpha}^{\dagger}c_{k\bar{\sigma},\alpha}d_{\sigma}|d_{\sigma}^{\dagger}}\rangle\rangle \approx \langle{c_{k^{'}\bar{\sigma},\alpha}^{\dagger}c_{k\bar{\sigma},\alpha}}\rangle \langle\langle{d_{\sigma}|d_{\sigma}^{\dagger}}\rangle\rangle
\end{align*}
Combining Eqns. (4)-(8) and taking $U\rightarrow\infty$ limit, the final expression for the single particle retarded Green's function for dot is given as
\begin{widetext}
\begin{equation} \label{eq:8}
G_{d\sigma}^{r}(\omega)=\langle\langle{d_{\sigma}|d_{\sigma}^{\dagger}}\rangle\rangle_{\omega}=\left[\cfrac{1-\langle{n_{\bar{\sigma}}}\rangle-\sum_{k,\alpha}V_{\alpha}\cfrac{\langle{d_{\bar{\sigma}}^{\dagger}c_{k^{'}\bar{\sigma},\alpha}}\rangle}{\omega-\epsilon_{k,\alpha}}}{\omega-\epsilon_d-\sum_{k,\alpha}\cfrac{V^{2}_{\alpha}}{\omega-\epsilon_{k,\alpha}}-\sum_{kk^{'},\alpha}V_{\alpha}^2\cfrac{\langle{c_{k^{'}\bar{\sigma},\alpha}^{\dagger}c_{k\bar{\sigma},\alpha}}\rangle}{\omega-\epsilon_{k,\alpha}}+\sum_{k,\alpha}\cfrac{V^{2}_{\alpha}}{\omega-\epsilon_{k,\alpha}}\sum_{k,\alpha}V_{\alpha}\cfrac{\langle{d_{\bar{\sigma}}^{\dagger}c_{k^{'}\bar{\sigma},\alpha}}\rangle}{\omega-\epsilon_{k,\alpha}}}\right]
\end{equation}\\
\end{widetext}
The expectation values $\langle{n}_{\bar{\sigma}}\rangle$, $\langle{d_{\bar{\sigma}}^{\dagger}c_{k^{'}\bar{\sigma},\alpha}}\rangle$ and $\langle{c_{k^{'}\bar{\sigma},\alpha}^{\dagger}c_{k\bar{\sigma},\alpha}}\rangle$ are calculated self-consistently using  spectral theorem:
\begin{align}
\label{eq:9}&\langle{n}_{\bar{\sigma}}\rangle =-\cfrac{1}{\pi}\int{f(\omega)Im\left\{\langle\langle{d_{\sigma}|d_{\sigma}^{\dagger}}\rangle\rangle\right\}d\omega}\\
\nonumber\\ 
\label{eq:10}&\langle{d_{\bar{\sigma}}^{\dagger}c_{k^{'}\bar{\sigma},\alpha}}\rangle =-\cfrac{1}{\pi}\int{f({\omega^{'}})Im\left\{\langle\langle{c_{k^{'}\bar{\sigma},\alpha}|d_{\bar{\sigma}}^{\dagger}}\rangle\rangle\right\}d{\omega^{'}}}\\
\nonumber\\
\label{eq:11}&\langle{c_{k^{'}\bar{\sigma},\alpha}^{\dagger}c_{k\bar{\sigma},\alpha}}\rangle =-\cfrac{1}{\pi}\int{f({\omega^{'}})Im\left\{\langle\langle{c_{k\bar{\sigma},\alpha}|c_{k^{'}\bar{\sigma},\alpha}^{\dagger}}\rangle\rangle\right\}d{\omega^{'}}}
\end{align}
with
\begin{align}
\label{eq:12}&\langle\langle{c_{k^{'}\bar{\sigma},\alpha}|d_{\bar{\sigma}}^{\dagger}}\rangle\rangle =\cfrac{V_{\alpha}}{\omega^{'}-\epsilon_{k,\alpha}}\langle\langle{d_{\sigma}|d_{\sigma}^{\dagger}}\rangle\rangle\\
\nonumber\\
&\nonumber\langle\langle{c_{k\bar{\sigma},\alpha}|c_{k^{'}\bar{\sigma},\alpha}^{\dagger}}\rangle\rangle =\cfrac{V^{2}_{\alpha}}{(\omega^{'}-\epsilon_{k,\alpha})(\omega^{'}-\epsilon_{k^{'},\alpha})}\langle\langle{d_{\sigma}|d_{\sigma}^{\dagger}}\rangle\rangle\\
\label{eq:13}&\hspace{6.5em}+\cfrac{\delta_{kk^{'}}}{{\omega^{'}}-\epsilon_{k,\alpha}}
\end{align}
where $\omega^{'}=\omega^{'}+i\delta$ is dummy variable and $\delta\rightarrow 0^+$.\\
The terms with summations over $k$ appearing in the expression of Green's function in Eqn. (9) can be simplified by replacing $\sum_k \rightarrow \int{\rho(\epsilon)d\epsilon }$ and then solving these expressions using the complex contour  integration in the flat wide band limit i.e.
\begin{align}
\label{eq:14}&\sum_{k,\alpha}\cfrac{V^{2}_{\alpha}}{\omega-\epsilon_{k,\alpha}} = \cfrac{1}{2}\sum_{\alpha}\int{\cfrac{\Gamma_{\alpha}}{\omega-\epsilon}}\,d\epsilon =-i\Gamma-ln\left|\cfrac{D-\omega}{D+\omega}\right|\\
\label{eq:15}&\sum_{k,\alpha}V_{\alpha}\cfrac{\langle{d_{\bar{\sigma}}^{\dagger}c_{k^{'}\bar{\sigma},\alpha}}\rangle}{\omega-\epsilon_{k,\alpha}} = \cfrac{\Gamma}{\pi}\int{\cfrac{f({\omega^{'}})G_{d\sigma}^{a}({\omega^{'}})}{\omega-\omega^{'}+i\delta}}d\omega^{'}\\
\nonumber\\
\label{eq:16}&\sum_{kk^{'},\alpha}V_{\alpha}^2\cfrac{\langle{c_{k^{'}\bar{\sigma},\alpha}^{\dagger}c_{k\bar{\sigma},\alpha}}\rangle}{\omega-\epsilon_{k,\alpha}} = \cfrac{\Gamma}{\pi}\int{\cfrac{f({\omega^{'}})\left(1+i\Gamma G_{d\sigma}^{a}({\omega^{'}})\right)}{\omega-\omega^{'}+i\delta}}d\omega^{'}
\end{align}
In above integral equations the advanced Green's function $G_{d\sigma}^{a}(\omega) = \left[G_{d\sigma}^{r}(\omega)\right]^{\ast}$ and Fermi-Dirac function $f(\omega)=\left[exp(\omega/k_BT)+1\right]^{-1}$. To obtain the expectation values $\langle{n}_{d\bar{\sigma}}\rangle$, $\langle{d_{\bar{\sigma}}^{\dagger}c_{k^{'}\bar{\sigma},\alpha}}\rangle$ and $\langle{c_{k^{'}\bar{\sigma},\alpha}^{\dagger}c_{k\bar{\sigma},\alpha}}\rangle$ and final Green's function, the set of Eqns. (9)-(12) are solved self-consistently.\\
To further simplify the calculation, the most commonly used approximation is the one worked out by Meir et. al. \cite{Meir1991}. In this approximation, the expectation values are treated non-self-consistently, following the lowest-order approximation i.e. $\langle{d_{\bar{\sigma}}^{\dagger}c_{k^{'}\bar{\sigma},\alpha}}\rangle\approx 0$ and $\langle{c_{k^{'}\bar{\sigma},\alpha}^{\dagger}c_{k\bar{\sigma},\alpha}}\rangle\approx f(\epsilon_{k,\alpha})\delta_{kk^{'}}$. In the next section, a comparison is made between the self-consistent and non-self-consistent solutions, and it is shown that the latter provides a good description of the situation for temperatures close to or above the Kondo temperature $T_K$. However, it is not a correct approach for temperatures much lower than $T_K$.\\
The spectral density on the quantum dot is given as $\rho_d(\omega)=-\cfrac{1}{\pi}\sum_{\sigma}\left[Im\left\{G_{d\sigma}^{r}(\omega)\right\}\right]$. In the linear response regime, i.e., for  small Voltage biasing ($\mu_L-\mu_R=e\delta V\rightarrow 0$) and small temperature gradients ($T_L-T_R=\delta\theta\rightarrow 0$) between the leads, the electrical current and heat current satisfy the Onsager relation\cite{Mahan2000},
\begin{equation}\label{eq:17}
  \begin{pmatrix}
 I_C \\
 \\
 J_Q \\
 \\
 \end{pmatrix}
 =
  \begin{pmatrix}
 e^2L_0 &\hspace{1.5em} \cfrac{e}{T}L_1 \\
 \\
 eL_1 &\hspace{1.5em} \cfrac{1}{T}L_2 \\
 \end{pmatrix}
 \\
  \begin{pmatrix}
 \delta V \\
 \\
  \delta\theta \\
  \\
 \end{pmatrix}
\end{equation}
with thermoelectric response functions
\begin{equation*}
\begin{aligned}
L_0 &= \cfrac{2}{h}\int{\left(\cfrac{-df(\omega)}{d\omega}\right) T(\omega)d\omega}\\
\\
L_1 &= \cfrac{2}{h}\int{\omega\left(\cfrac{-df(\omega)}{d\omega}\right) T(\omega)d\omega}\\
\\
L_2 &= \cfrac{2}{h}\int{\omega^2\left(\cfrac{-df(\omega)}{d\omega}\right) T(\omega)d\omega}
\end{aligned}
\end{equation*}
Here, $e$ and $h$ denote the magnitude of the electronic charge and Planck’s constant, respectively. $T(\omega)=\Gamma^2 |G_{d\sigma}^r(\omega)|^2$ is the tunnelling amplitude. The thermoelectric transport quantities (electrical conductance $G$, thermopower or Seeback coefficient $S$, and electronic contribution to thermal conductance $K$) are then obtained from Eqn. (18) as
\begin{align}
\label{eq:18}
G = & \lim_{\delta V \to 0} {\cfrac{I_C}{\delta V}}\biggr\rvert_{\delta\theta=0} = e^2L_0\\
\nonumber\\
\label{eq:19}
S = & \lim_{\delta\theta\to 0} {\cfrac{\delta V}{\delta\theta}}\biggr\rvert_{I_C=0} = -\cfrac{1}{eT}\frac{L_1}{L_0}\\
\nonumber\\
\label{eq:20}
K = & \lim_{\delta\theta \to 0} {\cfrac{J_Q}{\delta\theta}}\biggr\rvert_{I_C=0} = \cfrac{1}{T}\left[L_2-\cfrac{L_1^2}{L_0}\right]=\cfrac{1}{T}L_2-S^2GT
\end{align}
The performance of the thermoelectric particle-exchange heat engine in the linear response regime is determined by a dimensionless thermoelectric figure of merit $ZT$.
\begin{equation}\label{eq:21}
ZT = \cfrac{S^2GT}{K}
\end{equation}
Since we are interested in understanding the electronic thermal characteristics at low temperatures, we only take into account the thermal contribution from electrons while excluding the thermal contribution from the lattice or phonons, which is insignificant for the present case.
\section{Result and discussion}\label{sec:artwork}
Based on the equations derived in the preceding section, the numerical computations are performed in MATLAB. The self-consistent Eqns. (9)-(12) are solved iteratively for the expected values until convergence is reached. The resulting Green's function is then used to determine the spectral and thermoelectric transport quantities with $\Gamma_0$ (in meV) as the energy unit.\\
\begin{figure*}[!htb]
\includegraphics [width=0.65\hsize]{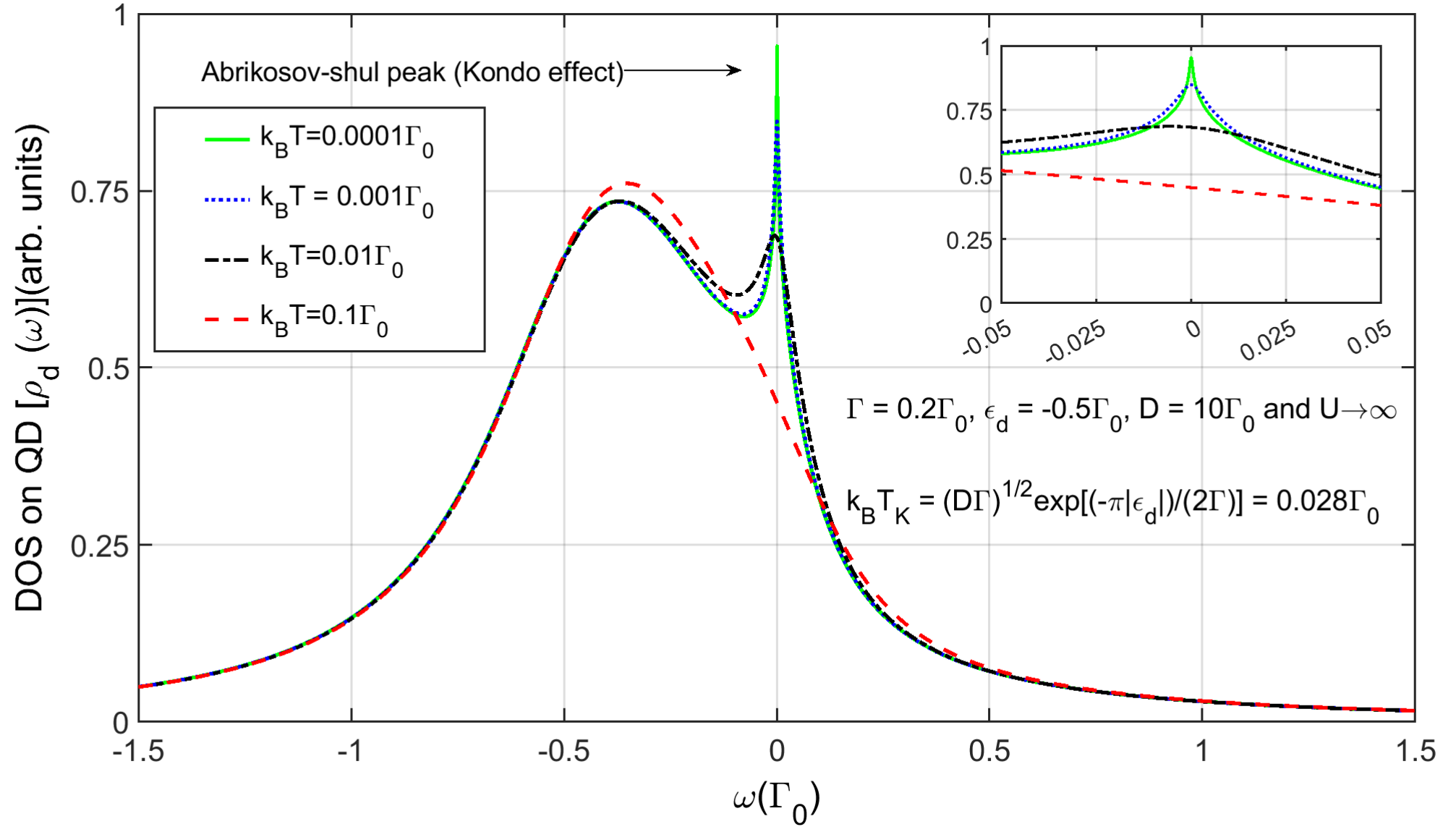}
\caption {Density of states $\rho_d(\omega)$ versus energy $\omega$ for different $k_BT$. The other parameters are: $\epsilon_d=-0.5\Gamma_0$, $\Gamma=0.2\Gamma_0$ and $D=10\Gamma_0$. The insets show a zoom near the region $\omega=0$}
\end{figure*}
Previous research works on quantum dot based systems has revealed that the density of states of the QD exhibits a characteristic Abrikhosov-shul peak at Fermi energy for $T\leq T_K$, as well as Hubbard peaks for entire temperature range. The Abrikhosov-shul peak is a signature of the Kondo effect and is a key feature in the characterization of the electronic properties of these systems at low temperature. In Fig. 1, we have plotted a variation of density of states with energy for different stages of temperatures. At very low temperatures, Kondo peak is clearly observed which suppresses as temperature increases. At certain temperature equivalent to Kondo temperature ($T_K$), the Kondo peak disappears indicates the dominating behavious of Coulomb blockade and thermal flactuations over Kondo effect. On the other hand, the Hubbard peaks, which represent a single electron and hole excitation, are unaffected by the thermal fluctuations.\\
\begin{figure*}[!htb]
\includegraphics [width=0.9\hsize]{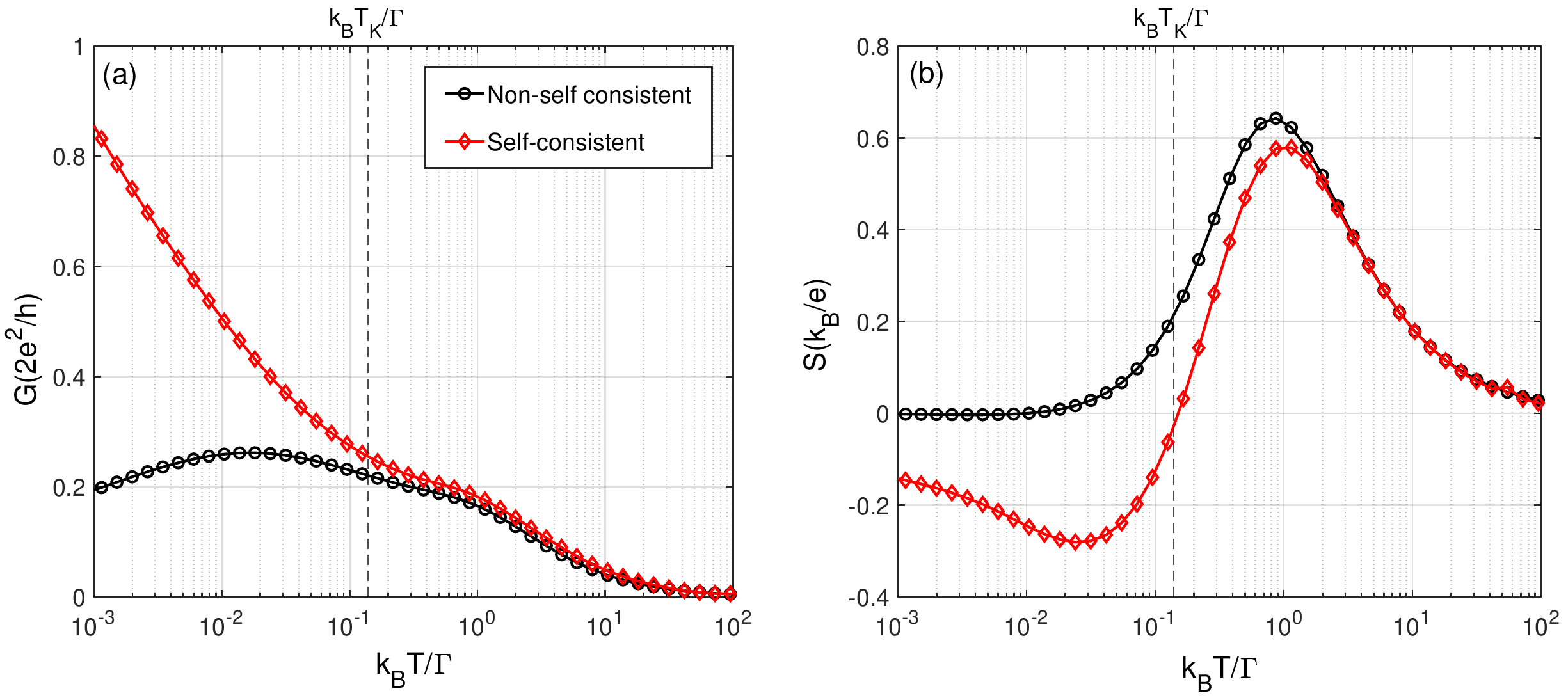}
\caption {(a) Linear electrical conductance $G$ and (b) linear thermopower $S$ versus thermal energy $k_BT$ for a strongly interactig QD ($U\rightarrow\infty$) using the non-self consistent\cite{Meir1991,Sun2001,Zhang2002,Martinek2003,Tolea2007,Mathe2020,Verma2023} and complete self-consistent approach\cite{Kashcheyevs2006,Wohlman2005,Fang2008} in EOM. The other parameters are same as in Fig. 1. The vertical dashed lines indicate the thermal energy corresponding to the Kondo temperature $T_K$ i.e., $k_BT_K=(D\Gamma)^{1/2}exp\left[{(-\pi|\epsilon_d|)/(2\Gamma)}\right]$.}
\end{figure*}
Fig. 2 depicts the variation of electrical conductivity $G$ and thermopower $S$ with temperature for self-consistent and non-self-consistent solutions. The electrical conductivity and thermopower are clearly independent of these solution processes at considerably higher temperatures ($T>T_K$, where $T_K$ is Kondo temperature). Therefore the non-self-consistent approximation previously used in various references\cite{Meir1991,Sun2001,Zhang2002,Martinek2003,Tolea2007,Mathe2020,Verma2023} describes the situation rather well for temperatures close to and above Kondo temperature. However, for $T<<T_K$, a significant difference in calculated quantities can be observed, which depends on the handling of Green's function and solving strategy. Hence, Green's function EOM technique can provide a qualitative description of thermoelectric transport across strongly interacting quantum dot systems in the Kondo regime, provided that the self-consistency conditions are fully taken into account. Various researchers have previously reported this temperature dependence of thermoelectric transport quantities using different theoretical techniques\cite{Boese2001,Krawiec2006,Krawiec2007,Dong2002,Kubala2006,Costi1994,Costi2010}, and a detailed discussion is provided in the next figure.\\
\begin{figure*}[!htb]
\centering
\includegraphics
  [width=0.95\hsize]
  {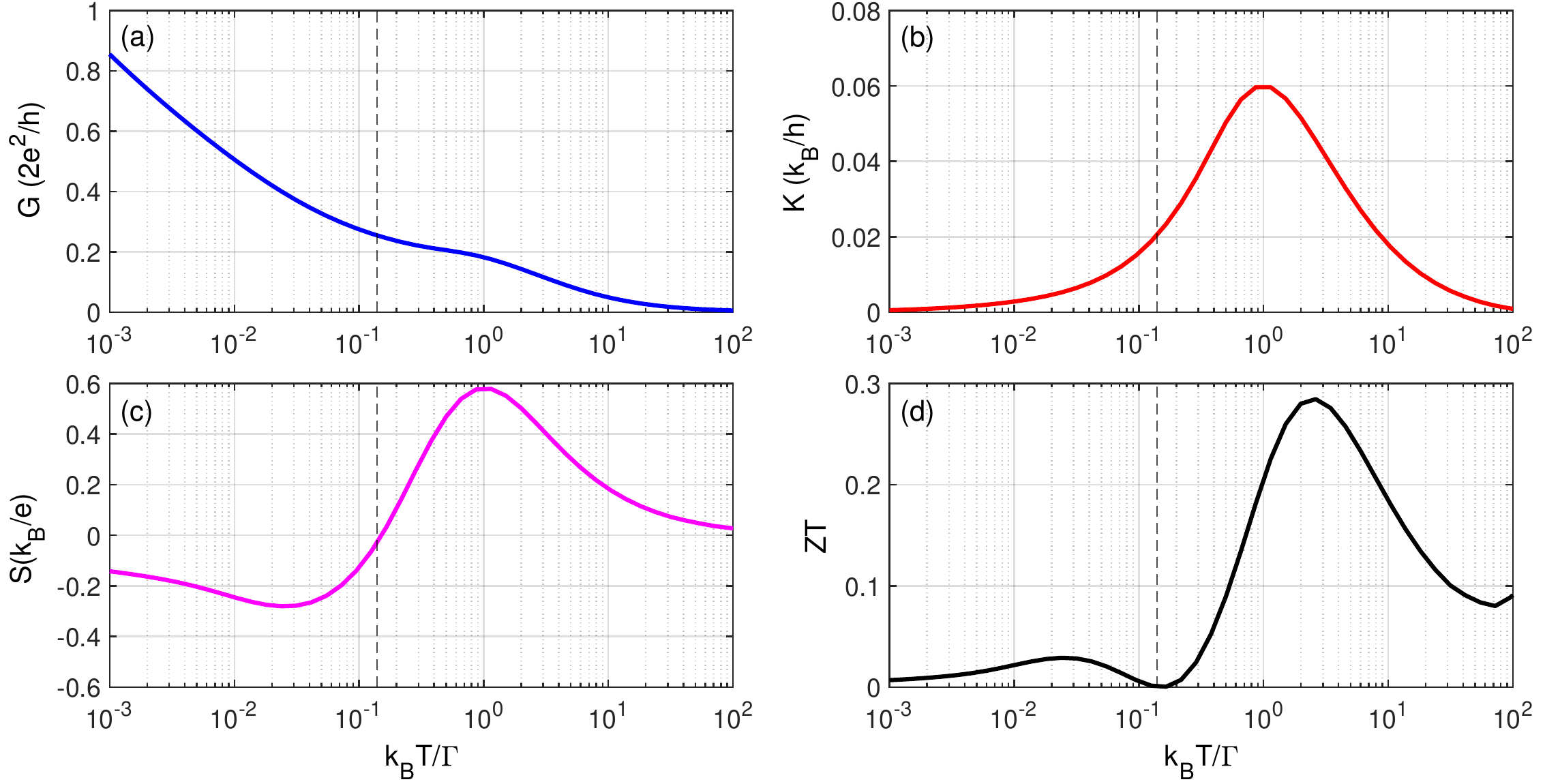}
\caption {Variation of the (a) electrical conductivity $G$, (b) electronic thermal conductivity $K$, (c) thermopower or Seebeck coefficient $S$ and, (d) figure of merit $ZT$ with the thermal energy $k_BT$ with the same parameters as in figure 1. The vertical dashed lines indicate the thermal energy corresponding to the Kondo temperature $T_K$.}
\end{figure*}
Following the above discussion, in Fig. 3, we have presented variations of electrical conductivity $G$, thermal conductivity $K$, thermopower $S$, and figure of merit $ZT$ with thermal energy range applicable to both the Kondo regime and Coulomb blockade regime. Fig. 3(a) shows the variation of electrical conductivity $G$ with temperature. At temperatures above the Kondo temperature, the electrical conductivity of the system is governed by the Coulomb blockade effect. In this regime, the electrons on the quantum dot tunnel one by one onto or off the dot due to their repulsive Coulomb interaction. When the energy required for electron tunnelling is much lower than the thermal energy, the sequential tunnelling process occurs, and electron transmission between the lead and the quantum dot occurs infrequently. Consequently, two successive tunneling events become incoherent, and the charge on the dot becomes nearly quantized. This results in a overall suppression of the electrical conductivity of the system. However, at temperatures below the Kondo temperature, the Kondo effect or higher order tunnelling processes begins to dominate, leading to an increase in the electrical conductivity of the system. The impurity spin in the quantum dot becomes screened by the conduction electrons in the surrounding leads, forming a Kondo singlet. This results in a reduction of the scattering of the electrons and an increase in the electrical conductivity of the system. At very low temperature $T\rightarrow{0}$, the electrical conductance approaches the unitary limit $2e^2/h$.\\ 
Fig. 3(b) shows that electronic contribution to the thermal conductance $K$ varies smoothly for a given temperature range and shows a broad maxima at $k_BT\approx\Gamma$ due to single particle excitation. The thermal conductance $K$ exhibits no unusual behavior around Kondo temperature $T_K$; thus, the Kondo state has no effect on $K$.\\
Fig. 3(c) illustrates that the linear thermopower $S$ changes sign close to the Kondo temperature ($k_BT_K\approx 0.028\Gamma_0$ in present case as indicated by a vertical black dashed line). This change in the sign of $S$ at $T_K$ is due to the change in the slope of the spectral density caused by the formation of the Abrikosov-shul peak at the Fermi level. If the QD energy level is below the Fermi level i.e $\epsilon_d<0$, then the transport at low temperature ($T<T_K$) is caused by electrons, whereas the transport at high temperature ($T>T_K$) is caused by holes.  Furthermore, $S$ exhibits a broad maximum at $k_BT\approx\Gamma$ due to single particle excitation. For $T\rightarrow{0}$, the recovery of Wiedemann-Franz law causes vanishing thermopower.\\
The thermoelectric figure of merit $ZT$, shown in Fig. 3(d), measures the system's utility for power generators or heat engines.  It is evident that $ZT$ vanishes at $k_BT_K\approx 0.028\Gamma_0$ and a significant figure of merit is observed in the Coulomb blockade regime when the Wiedemann-Franz law is violated. In the Kondo regime, $ZT$ shows small contribution due the low-temperature Kondo enhancement of the thermopower.\\
\begin{figure}[hbt!]
\centering
\includegraphics
  [width=1.0\hsize]
  {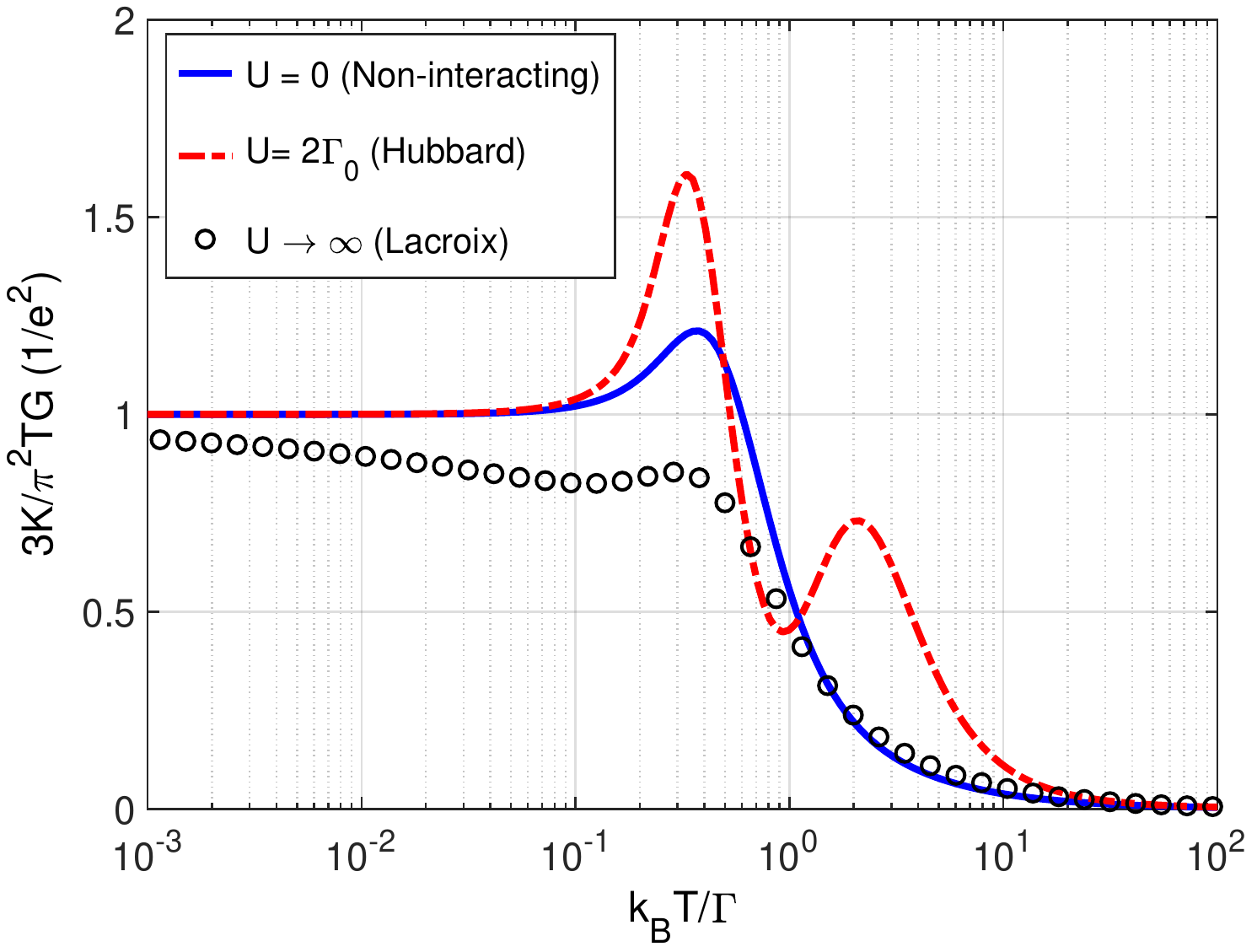}
\caption {Wiedemann-Franz ratio versus thermal energy for three different strengths of on-dot Coulomb interaction. The other parameters are same as in Fig. 1.}
\label{fig:two-columns-figure}
\end{figure}
Fig. 4 shows the Wiedemann-Franz ratio versus thermal energy obtained for our considered system for three different strengths of on-dot Coulomb repulsion, namely non-interacting case $U=0$, Hubbard mean field  $U=2\Gamma_0$ and strong interaction $U\rightarrow\infty$. The Wiedemann-Franz law a is fully obeyed at low temperatures ($ k_BT<<\Gamma$) for non-interacting and mean field correlated QD, showing that similar tunneling processes are responsible for charge and heat transport.  At relatively high temperatures, transport is dominated by sequential tunnelling events, resulting in a suppression of thermal conductance compared to electrical conductance. This suppression leads to the complete breakdown of the Wiedemann-Franz law, irrespective of the strength of the Coulomb interaction. This further indicates that the violation of the Wiedemann-Franz is mainly caused by the narrow Dirac-delta-shaped transmission function rather than the Coulomb correlation. For $U\rightarrow\infty$ the Kondo state creates an effective Fermi liquid state at low temperatures, and there is a substantial recovery in Wiedemann-Franz law.\\
These variations in linear thermoelectric quantities with thermal energy derived using Green's function equation of motion technique within Lacroix approximation exhibit qualitative agreement with previous theoretical works\cite{Boese2001,Dong2002,Kubala2006,Kubala2008,Krawiec2006,Krawiec2007,Franco2008,Costi1994,Costi2010,Bas2012,Weymann2013,Quintero2021}. It is also important to note that the approximations used in the present study are quantitatively accurate, far from the Kondo regime, and can only capture qualitative aspects of thermoelectric transport in the Kondo regime\cite{Kashcheyevs2006}.
\section{Conclusion}
In conclusion, to study the thermoelectric characteristics of a strongly interacting quantum dot coupled to metallic leads, we analyze the single impurity Anderson model using the equation of motion technique within the Lacroix approximation.  Through a comprehensive analysis of the linear thermoelectric properties, we demonstrate the necessity of a complete self-consistent treatment of the Green's function for a qualitative description of thermoelectric transport in the Kondo regime. The variations of electrical conductivity, thermal conductivity, thermopower, and figure of merit with thermal energy are examined. At low temperatures, the Kondo effect dominates, leading to increased electrical conductivity due to the screening of the impurity spin by conduction electrons. On the other hand, at higher temperatures, the Coulomb blockade effect governed the transport, resulting in the suppression of electrical conductivity. The thermal conductivity exhibits no unusual behavior around the Kondo temperature, while the thermopower changed sign close to the Kondo temperature. The figure of merit shows a significant value in the Coulomb blockade regime when the Wiedemann-Franz law is violated, while a small contribution is observed in the Kondo regime due to the Kondo enhancement of thermopower. These results are in qualitative agreement with other theoretical methods. We hope that the present self-consistent calculation of Green's function and results will be useful for future research into the thermoelectric characteristics of quantum dot-based novel hybrid systems.
\section*{Acknowledgements}
Sachin Verma is presently a research scholar at the department of physics IIT Roorkee and is highly thankful to the Ministry of Education (MoE), India, for providing financial support in the form of a Ph.D. fellowship.

\end{document}